\documentclass[final,authoryear,3p,times]{elsarticle}

\usepackage{graphics}

\journal{New Astronomy}

\begin{document}

\begin{frontmatter}



\title{Thermal characteristics of a classical solar telescope primary mirror}


\author{Ravinder K Banyal}
\ead{banyal@iiap.res.in}
\author{and B. Ravindra}
\address{Indian Institute of Astrophysics, Bangalore 560034, INDIA}

\begin{abstract}
We present a detailed thermal and structural analysis of a 2m class solar telescope mirror which is subjected to a varying heat load at an observatory site. A 3-dimensional heat transfer model of the mirror takes into account the heating caused by a smooth and gradual increase of the solar flux during the day-time observations and cooling resulting from the exponentially decaying ambient temperature at night. The thermal and structural response of two competing materials for optical telescopes, namely Silicon Carbide -best known for excellent heat conductivity and Zerodur -preferred for its extremely low coefficient of thermal expansion, is investigated in detail. The insight gained from these simulations will provide a valuable input for devising an efficient and stable thermal control system for the primary mirror.

\end{abstract}


\end{frontmatter}


\section{Introduction}\label{intro}
At the beginning of 17th century, Galileo Galilei turned his small telescope towards the sun and noticed the dark regions on sun's surface now called sunspots. The larger telescopes built subsequently revealed that sunspots have different structures like dark region called umbra and less dark with filamentary structures called penumbra.  As the technology improved in 20th century, bigger telescopes were built to observe the characteristic features which include the  light bridge across sunspots \citep{vaz73}, moving magnetic features around sunspots \citep{har73}, supergranulation network and its elements \citep{sim64} etc. Most of these observations have been done using about 50 cm class telescopes or smaller. From the theoretical calculations, the size of the elementary building blocks (size of the flux tube) of magnetic structures, for example, is estimated to be about 70 km. A large aperture telescope with diameter of 2m or more is therefore, necessary to resolve  important and intriguing features of the solar atmosphere at smaller scales.  Additionally, the photon starved polarization measurements, crucial to estimate the magnetic fields distribution and strength in the active regions on the sun, are hard to come by from small telescopes.

The next generation, large aperture (2m and above) solar telescopes equipped with adaptive optics and supported by
advanced backend instruments, will be able to probe the solar atmosphere at unprecedented details and thus help solving
many outstanding problems in solar physics. Some of the notable solar facilities planned for the future are:
a 4m advanced technology solar telescope (ATST) to be build by National Solar Observatory \citep{kei09}, National Large Solar Telescope (NLST) proposed by Indian Institute of Astrophysics, INDIA \citep{sin08,has10}, and a soon to be commissioned 1.5 m Gregor telescope by a consortia led Kiepenheuer-Institute, Germany \citep{pet06}.

The design and construction of a solar telescope is considerably different from a typical night-time telescope. Among others, the most formidable task is  to build an efficient temperature control system capable of handling the large amount of heat generated by the intense solar radiation. The immense difficulty of the heat management problem has been a major bottleneck in constructing large aperture primary mirror for solar studies. Evidently, by the time largest solar telescope (4m ATST) starts its operation, the night-time astronomy would have leap-frogged to next generation of 20-40m class telescopes.

Besides the direct heating caused by light absorption, the diurnal temperature variations of ambient air are also crucial in driving the thermal response of the telescope mirror. The rise or fall in mirror temperature with respect to ambient, adversely affect the imaging performance of the telescope in two major ways. First, when the mirror is at different temperature than the surrounding, a thin layer with a temperature gradient is formed closed to the  mirror surface and air.  The temperature fluctuations in the surface layer lead to the variations in the refractive index of the air that produce wavefront aberrations \citep{low79,sal92,bar93,woo95}. The final image is blurred as the telescope beam passes through this layer twice. The second detrimental effects of heating or cooling is the structural deformations of the mirror arising from thermally induced stress and temperature inhomogeneities within the substrate volume \citep{pea87,spi67,bur99}. Any deviation from the nominal temperature distorts the mirror surface from its desired shape. Thermally induced surface imperfections therefore, deteriorate the final image quality of the telescope.

To minimize the effect of mirror seeing, the temperature difference between the ambient air and the mirror surface should be maintained within a very narrow ($<\pm2^\circ$C) range \citep{emd04,dal04}. At the same time, the temperature difference across the reflecting face of the mirror should not exceed $\pm0.5^\circ \textrm{C}$. For night-time astronomy, the temperature of the mirror and telescope dome can be effectively controlled by air conditioning system during the day \citep{miy103}. Blowing cool air over the mirror surface is helpful in mitigating the mirror seeing effects. Forced air flow breaks the thermal plumes closed to the mirror surface and thus homogenize the temperature gradients which otherwise drive the refractive index fluctuations leading to image degradation \citep{low79}. The induced air flow also improves the convective heat transfer rate and thereby reducing the temperature difference between the mirror material and the surrounding air. For effective cooling, the air temperature and speed have to be regulated according to the temperature profile of the mirror surface. However, the accurate measurements of the temperature of the mirror require an array of temperature sensors placed close to the reflecting face of the mirror which is not just difficult but impractical during the scientific observations. In such a case, numerical simulations are the only viable solution to predict the temperature profile of the mirror surface.

In another approach, a `cool reservoir' is created by lowering the the mirror temperature well below the temperature expected for the night observations. A near thermal equilibrium with ambient is accomplished in a resistive heating by passing electrical current through reflective coating on the front surface of the mirror \citep{boh00,gre94}. This approach does not quite suit the solar observations as the primary mirror is directly heated by sun's radiation. The effect of mirror seeing is even more severe if the same telescope is later switched to night-time observations. To avoid large heat loads and thermal gradients, ideally a substrate material should have low heat capacity  and high thermal conductivity.

Among traditional materials used in primary mirror substrates, ultra low expansion (ULE) glass ceramics such as zerodur and fused silica  have gained wide acceptance among astronomical community \citep{doh_1_07}. The semiconducting SiC is another important material that is beginning to find a niche in astronomy \citep{ben93,rit96,pet94,tsu05,mat08}. In addition to its low thermal expansion coefficient, high hardness and rigidity, SiC has exceptionally high thermal conductivity which is about two order of magnitude higher than the other materials. For large size telescopes, the lightweighted mirror geometry is preferred  to reduce the overall weight and thermal inertia of the system. The reduced mass facilitates efficient and faster cooling necessary for the mirror to reach the thermal equilibrium with surroundings quickly. However,  lightweighting is a complex, time consuming and often an expensive process. With a good thermal control system, a classical primary mirror of diameter not exceeding 2m could possibly be used without lightweighting.

For a given shape and geometry of the mirror, the analytical solutions become intractable. Therefore, in traditional engineering practice, the thermal response of such a mirror blank is studied by finite element methods (FEM) \citep{cho95,dal04}. The heat transfer problem described by a diffusion-type partial differential equation, is numerically solved to obtain a steady-state solutions for a constant temperature typically found at telescope location. In addition, the mirror heating is modeled assuming fixed heat flux and fixed ambient air temperature which serves as the boundary conditions.  The time scale for a mirror to reach a thermal equilibrium is of the order of few days while the ambient air temperature changes significantly in course of the day \citep{boh00}. This means, the usual mirror substrate with large heat capacity can never reach the thermal equilibrium with ambient unless some external cooling is used. The static analysis is not enough to determine the level of accuracy and other parameters of interest (e.g. level of wind flushing or ventilation points etc) necessary for designing an efficient thermal control system.  A complete time dependent solution of 3D heat transfer problem under the varying solar flux and ambient temperature is therefore necessary to evaluate the mirror performance, a problem which we address in the present paper.

In this paper, we have solved the 3D heat transfer and structural model that takes into account the  ambient thermal conditions existing at the telescope site. The location dependent solar flux and ambient heating model is incorporated into commercially available finite element analysis software to investigate the time dependent thermal and structural response of 2m class primary mirror. These numerical simulations are carried out at four different ambient temperature range, representative of mild to extreme thermal variations that may occur at different observatory locations around the world.

\subsection{Formulation of heat transfer problem}
The general heat flow problem involves all three modes of heat transfer, namely conduction, convection, and radiation. The reflecting face coated with a thin metallic layer absorbs about 10-15\% of the total solar flux incident on the mirror surface.  The heat generated as a result of absorption of optical energy is diffused to other parts of the mirror via conduction and  partly lost to the surroundings via convective and radiative processes. The temperature of the mirror rises when the heat produced by light absorption exceeds the overall heat loss to surroundings. The general heat transfer problem can be cast into the partial differential equation of the form \citep{inc01}:
\begin{equation}\label{eq1}
    \rho\;C_{\textrm{p}}\frac{\partial T}{\partial t}-\nabla\cdot(k\cdot\nabla T)+\rho\;C_\textrm{p}\;\textbf{\textrm{u}}\nabla\cdot T = Q
\end{equation}
where $\rho$, $T$, $k$ and  $C_\textrm{p}$ are density, temperature, thermal conductivity and heat capacity of the material. The terms $\textbf{\textrm{u}}$ and $Q$ represent the velocity field and heat source, respectively. Most of the material properties are temperature dependent. The boundary condition for the heat flux to be maintained at the surface can be written as:
\begin{equation}\label{eq1}
    \textbf{\textrm{n}}\cdot (k \nabla T)=q_{0} + h(T_{\textrm{amb}}-T) + \epsilon \sigma (T_{s}^{4}-T^4)
\end{equation}

where $q_{0}$ is the inward heat flux at the mirror surface, $T_{\textrm{amb}}$ is the ambient air temperature, $T_{s}$ is the effective radiation temperature of the surroundings, $\epsilon$ is surface emissivity, $\sigma$ is the Stefan-Boltzmann constant and $h$ is the heat transfer coefficient indicating the rate at which heat is exchanged between the mirror and the surroundings. The 2nd and the 3rd terms in Eq.(2) are statements of Newton's law of cooling and Stefan-Boltzmann's law, accounting for the heat loss by free convection and radiation, respectively.

\subsection{Modeling the diurnal temperature cycle}
The diurnal and annual temperature variations is one of the important criteria for selecting a suitable observatory site. It also plays a vital role in determining the degrading effects of  atmospheric turbulence, mirror seeing and structural stability of the telescope. The earth's surface is mainly heated by the irradiated solar energy during the day. The sun's light also contains the spectral signatures of both solar and terrestrial atmospheres \citep{wal94}. The amount and duration of the solar radiation $I(\textrm{Wm}^{-2})$ reaching the earth's surface is given by \citep{iqb83}
\begin{equation}\label{eq3}
    I=I_{0}\,\cos(Z)
\end{equation}
where $I_{0}$ is the flux amplitude when sun is at zenith, and $Z$ is the solar zenith angle which is given by
\begin{equation}\label{eqs4}
    Z=\arccos(\sin\delta\sin\phi + \cos\delta\cos\phi\cos H)
\end{equation} where $\phi$ is latitude of the place, $\delta$ is declination angle of the sun, $H=(\pi/12)\;t$ is the solar hour angle with respect to noon and $t$ is the local time.

The exact nature of daily temperature variations at a given location, however, depends on the local weather conditions (pressure, wind speed, humidity etc), surface topography, soil type, vegetation, and the presence of water body etc. \citet{got01} proposed a simple physics-based model to describe the thermal heating of the Earth's surface in cloud free conditions. According to this model, the day-time rise in temperature  $T_{1}(t)$ due to the sun  and the night-time cooling $T_{2}(t)$ can be described by
 \begin{eqnarray}
    T_{1}(t)\!&=&\! s\,T_{0}+s\;T_{a} \cos\left[\frac{\pi}{\omega}(t-\tau_m)\right]\qquad\qquad\qquad\qquad\textrm{for}\quad \;\;t<t_s \\
   T_{2}(t)\!&=&\!s(T_{0}+\delta T) + s\left\{T_{a} \cos\left[\frac{\pi}{\omega}(t-\tau_m)\right]-\delta T \right\}  \textrm{e}^{-\frac{t-t_s}{\kappa}}
    \;\;\textrm{for} \;\;t \geq t_s \;\;\;     \label{eqs56}
 \end{eqnarray}

 For heating part, the choice of harmonic term in Eq.(5) is based on the solution of equation of thermal diffusion \citep{car59}, while the exponential term in Eq.(6) accounts for the temperature fall in accordance to the Newton's law of cooling. The constant $\tau_m$ is included to account for the thermal time lag between the peak solar flux and the peak ambient temperature during the day. The meaning and values of various parameters in Eq.(5) and Eq.(6) are listed in Table \ref{tab1}.
 \begin{table}[h]
  \centering \caption{Model parameters and the values for Eqs.\ref{eqs56}}\label{tab1}
  \begin{tabular}{lll}
       \hline
   Parameter & Meaning & Values \\
   \hline
    $T_{a}$ & temperature amplitude & 6.8 $^{\circ}$C  \\
    $T_{0}$ & residual temperature  & 3.4 $^{\circ}$C \\
    $\delta T$ & $T_{0}-T(t\rightarrow\infty)$  & -3.5 $^{\circ}$C\\
    $\omega$ & half period oscillation of cosine term  & 12 hrs\\
    $\kappa$ & attenuation constant  & 3.5 hrs \\
    $\tau_{m}$ & time when temperature maxima is reached  & 14:00 hrs\\
    $t_{s}$ & time when temperature attenuation begins & 18:00 hrs\\
    $s$ & scaling factor & 1.0 \\
       \hline \hline
  \end{tabular}
    \end{table}
  These values are arbitrarily chosen to simulate a typical temperature profile for $0-10^\circ$C ambient temperature variations. However, a location specific parameters can be obtained by model fitting the temperature data from the observatory site. The Eq.(3)-Eq.(6) prescribe the necessary boundary conditions and initial conditions in our heat transfer model. The FEM model, thus, has all the necessary input for a realistic thermal analysis of the mirror.

\subsection{The FEM simulations}
\begin{figure}[h]
\centerline{\includegraphics[scale=0.3]{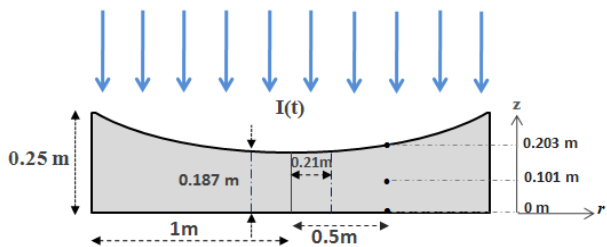}}
  \caption{Mirror geometry for heat transfer model. Three reference points are shown at radial distance $r=0.5$~m and different axial positions $z$ along the mirror thickness.   }\label{fig1}
\end{figure}

 For these studies we have chosen a Gregorian type primary mirror design which is most suitable for the solar observations. The Gregorian configuration allows the heat stop to be placed at the prime focus which is located before the secondary mirror \citep{von09}. This position is convenient to effectively block the unwanted radiation and prevent the subsequent heating of the secondary mirror and other optical components in the beam path. We chose a reflecting primary mirror with parabolic shape defined by z~$=r^2/4f$. The mirror diameter ($2r$) is 2m, focal length $f$ is 4m and radius of the central hole is 0.21m.  The edge-on geometry of the solid mirror is shown in Fig.~\ref{fig1}. This configuration is similar to the one under consideration for a 2m class NLST mirror \citep{has10,sin08}. The mirror is primarily designed for high resolution solar observations, but may partly be adopted for some limited but valuable scientific observations at night.

We used COMSOL multiphysics\footnote{COMSOL Multiphysics Pvt. Ltd., \texttt{http://www.comsol.com/}} as our FEM tool to study the thermal and structural characteristics of the solid mirrors. A free mesh containing over 34000 tetrahedral elements was created on the 3D geometry of the mirror. The reflecting surface of the mirror was modeled with a 100 nm thick aluminium coating layer. The thin aluminium  layer requires a much finer mesh which usually take longer computation time. Therefore, we have used the highly conductive layer approximation which does not require the resolved mesh for thin aluminium  layer. This approximation is valid since the thermal conductivity of aluminum is higher than the substrate material (see Table \ref{tab2} for comparison) and temperature across the aluminium layer thickness is nearly constant. In other words, any temperature gradient in the $l=100$~nm thick aluminum layer gets balanced over a minuscule time lag $\tau\sim l^2/(k/\rho C_\textrm{p})$.

As a generic case, the solid telescope mirror in the FEM model is exposed to solar radiation continuously between the sunrise (6:00 hrs) and sunset (18:00 hrs). Light rays are assumed to fall normally on the reflecting surface of the mirror. The peak solar flux reaching the earth surface at noon is assumed to be $1000\;\textrm{Wm}^{-2}$.  The thermal time lag between the peak solar flux and ambient air temperature is taken to be about 2 hours. Which means $T_{\textrm{amb}}$ continues to rise well past local solar noon, until 14:00 hrs. Eq. (3) is the source of inward heat flux generated at the front surface of the mirror during day-time heating $t<t_{s}$.  We have considered free convective heat loss from the mirror boundaries (front, back and sides) having heat transfer coefficient  $h=5\textrm{Wm}^{-2}$$\,^\circ\textrm{C}^{-1}$. The relevant thermal and mechanical properties of the mirror materials  are listed in Table \ref{tab2} \citep{lie08,pie02,ahm97}. The FEM model also includes their temperature dependence.

\begin{table}[!h]{\small
 \caption{Physical properties of materials at the room temperature.}\label{tab2}}
  \centering
         \begin{tabular}{|l|l l l|l|}
         \hline
         Property         &  \multicolumn{3}{c|}{Value}  & Units \\
          & SiC & Zerodur & Al &\\ \hline
         Thermal conductivity ($k$)    & 118  & 1.46  & 237    &  $\textrm{WK}^{-1} \textrm{m}^{-1}$ \\
         Specific heat capacity ($C_{\textrm{p}}$)  & 672  & 821   & 910    &  $\textrm{Jkg}^{-1}\textrm{K}^{-1}$ \\
         CTE ($\alpha$)                & 3.7  &-0.08  &  25    &  $10^{-6}$ $\textrm{K}^{-1}$ \\
         Material density ($\rho$)     & 3200 & 2530  & 2700   & kgm$^{-3}$ \\
         Young's modulus ($E$)         & 414  & 90.3  &  70    & $10^{9}$ Pa \\
         Poisson's ratio ($\nu$)       & 0.14 & 0.24  & 0.33   & --\\
         Emissivity ($\epsilon$)       & 0.9  & 0.9   & 0.05   &--\\
         \hline \hline
         \end{tabular}

\end{table}
The finite element structural model was used to predict the temperature induced mechanical deformations of the mirror blank. The mirror substrate will expand or contract during the heating and cooling cycles. Thermo-elastic response of the mirror can be examined at any desired instant by mapping the corresponding temperature patterns from the thermal model to structural model. The structural FEM model solves for material deformation for 3D geometry of the mirror. A fixed boundary condition was applied to the bottom face of the mirror to avoid the rigid body motion. The surface nodal displacement was computed for the remaining boundaries that were left free of external constrains.  No smoothing, transformation or interpolation of the thermal data was necessary as both thermal and structural models shared a common mesh on COMSOL software. Simulations were also performed at other ambient temperature range, $0\!-\!5^\circ$C, $0\!-\!20^\circ$C and $-10\!-\!0^\circ$C which correspond to the scaling parameter $s=0.5$, $s=2$ and $s=-1$, respectively, in Eq.(5) and Eq.(6).

\section{Results and discussions}
The 3D transient heat transfer model for Zerodur and SiC mirror was solved in two separate stages. The first stage run computes the solution for the  day-time ($6\!:\!00\leq\! t<18\!:\!00$) heating and second stage  run gives the solution for the night-time ($18:00\!\leq\!t<6\!:\!00$) mirror cooling. For each incremental time step ($\Delta T$=100s), the ambient temperature input was computed from Eq.(5) and Eq.(6). Shown in Fig. 2(a)  is the temperature evolution of three reference points located on (I) the front surface of the mirror ($r\!=\!0.5$m, $\!z=\!0.203$m), (II) the mid-plane of the mirror ($r\!=\!0.5$m, $z\!=\!0.101$m) and (III) the bottom of the mirror substrate ($r\!=\!0.5$m, $z\!=\!0.0$m). For comparison, the model input for the solar flux and $0\!-\!10^\circ$C ambient temperature curves are also drawn in Fig. 2(a). The grey dotted vertical line demarcates the heating ($t\!<\!t_s$) and cooling ($t\!\geq\!t_s$) regions.
\subsection{Code initialization}
Initially at $t=0$, the mirror and the ambient temperature was assumed to be 0$^\circ$C. Variations in ambient temperature and Zerodur mirror for the first 24 hrs simulation run are shown in Fig. 2(a). As the day progresses, the different parts of the mirror substrate exhibit dissimilar thermal lags in response to varying solar flux and ambient temperature. Also, for a significant part of the day, no part of the mirror exceeds the temperature of the ambient air. The hot ambient  also contribute to the mirror heating besides the apparent heating caused by the light induced absorption at the front surface. This is evident from the reference point at the mirror bottom ($z\!=\!0.0$m)  which remains at higher temperature \emph{w.r.t.} the midplane point ($z\!=\!0.101$m) until late afternoon hours when the temperature of the mirror interior is raised by the heat conduction from the front surface. After the sunset, the ambient air cools at a relatively faster rate.
\begin{figure}[h]
\centerline{\includegraphics[scale=0.25]{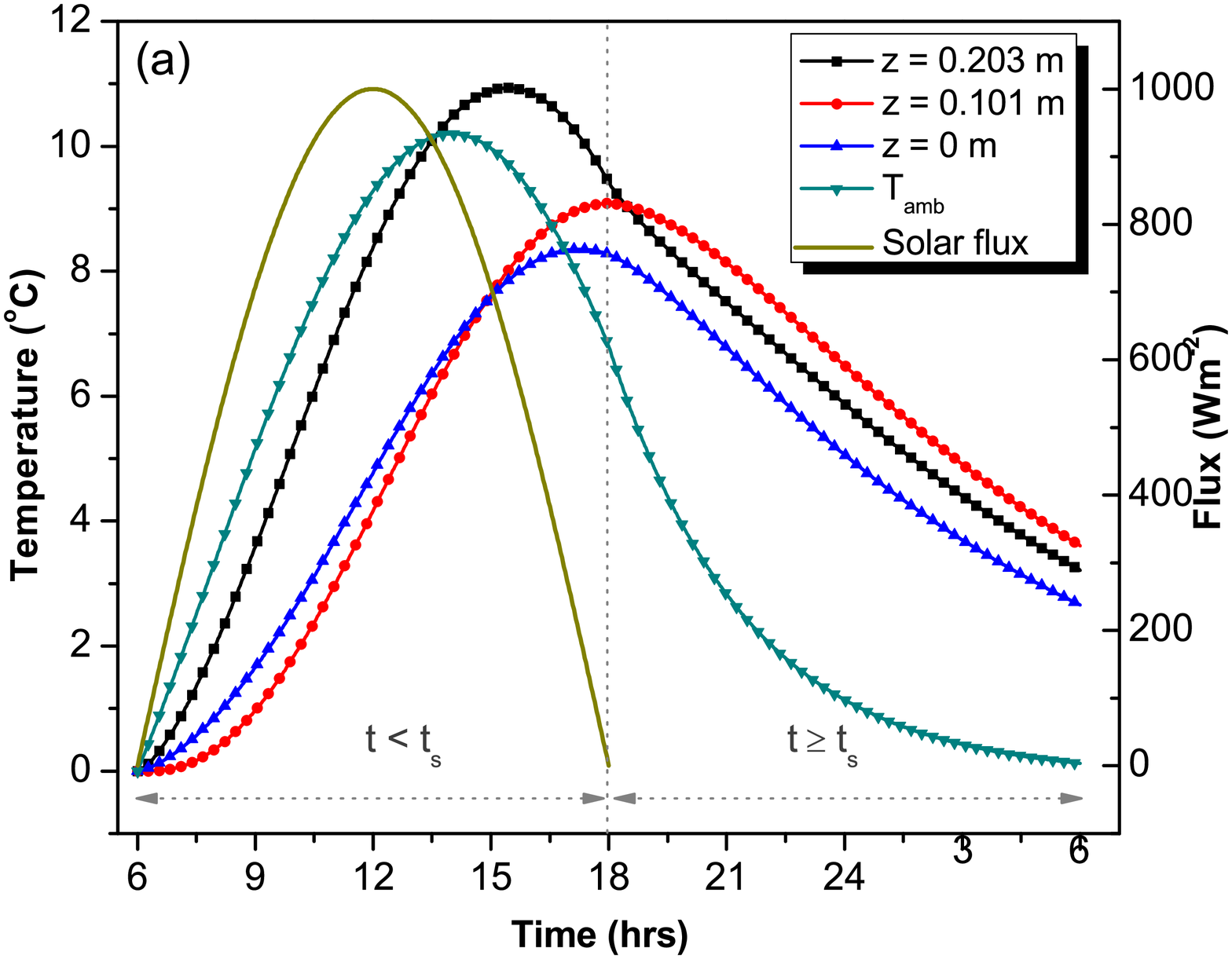} \includegraphics[scale=0.24]{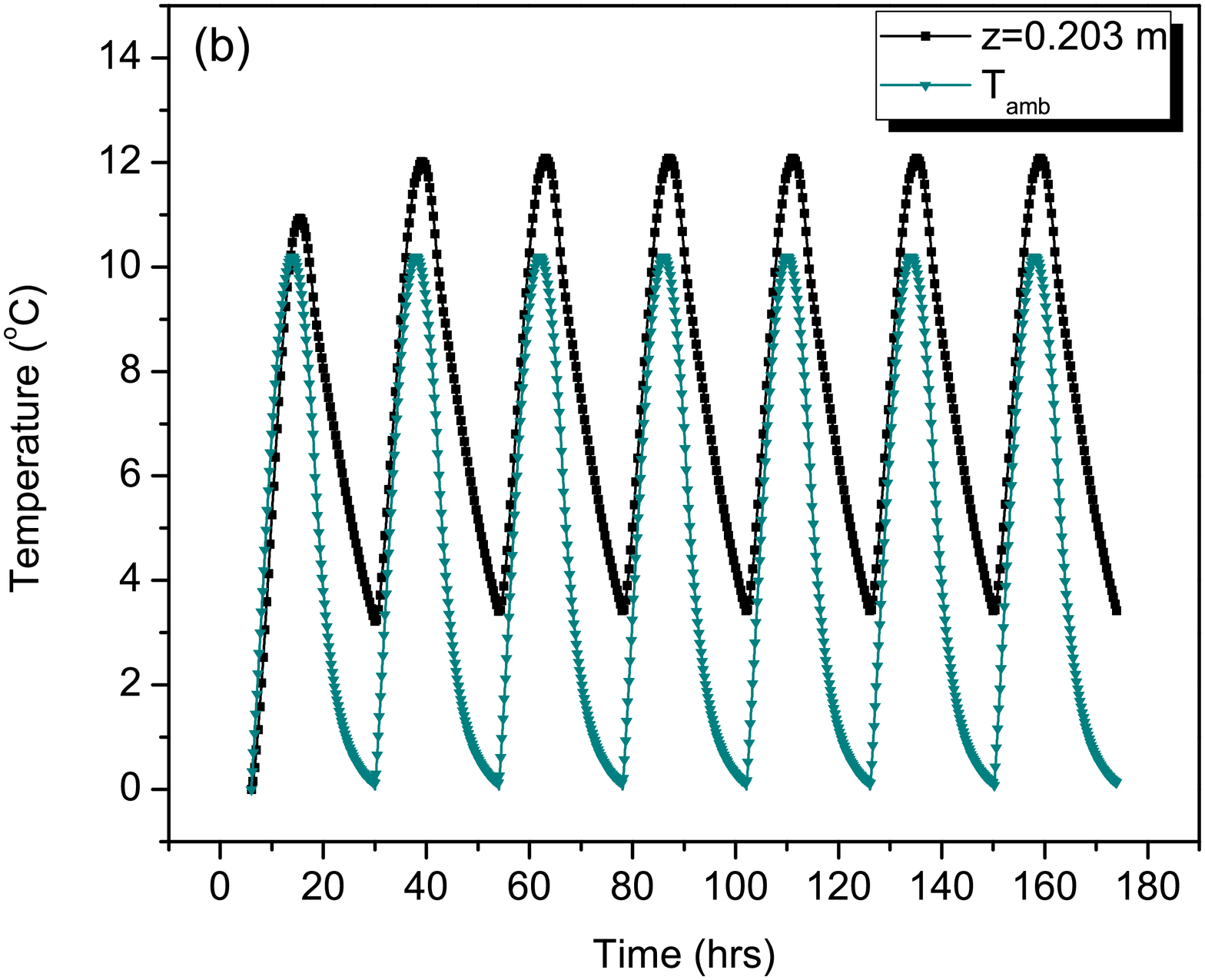}}
  \caption{Variations in ambient temperature and Zerodur mirror with time. (a) Simulation run for the first day. The day-time Solar flux curve is included to show the thermal lag between mirror temperature and ambient temperature. The scale for the solar flux is shown to the right. (b) Simulation run for the seven days observing cycle.}\label{fig}
\end{figure}

The slow fall in mirror temperature due to high thermal inertia (the mass $M$ and total heat capacity $MC_{p}$ of the Zerodur and SiC mirror being $\sim1672$~kg and $\sim2115$~kg, and $1.37 \times 10^6\;\textrm{JK}^{-1}$ and $1.42 \times 10^6\;\textrm{JK}^{-1}$ respectively), cannot keep up with the rapid fall in ambient air temperature.  The mirror and ambient temperatures differs by about 2-4$^\circ$C towards the end of first day simulation run. This has direct implications in the subsequent simulation for the next day observation cycle. The initial temperature conditions for the mirror in the next-day run would now be considerably different (2-4$^\circ$C higher than surroundings) than what they were for the start of the first day of observation. Fig. 2(b) shows the simulation run for seven days cycle. For clarity, the temperature curves for the front surface reference point ($z\!=\!0.203$m) and the ambient $T_{\textrm{amb}}$ alone are shown. Since the mirror does not cool to the surrounding temperature during night, there is an overall temperature increase of about $1^{\circ}$C above the ambient.  Except for the first-day simulation run, the thermal response of the mirror is identical for all other days. Mirror however, never attains thermal equilibrium with the ambient. From computing view point, the first-day simulation is more like a code initialization phase meant to produce a definitive starting temperature within the mirror for subsequent simulations. The code initialization is implicit in all subsequent results presented for Zerodur and SiC in remaining part of the paper.
\begin{figure}[h]
\centerline{\includegraphics[scale=0.3]{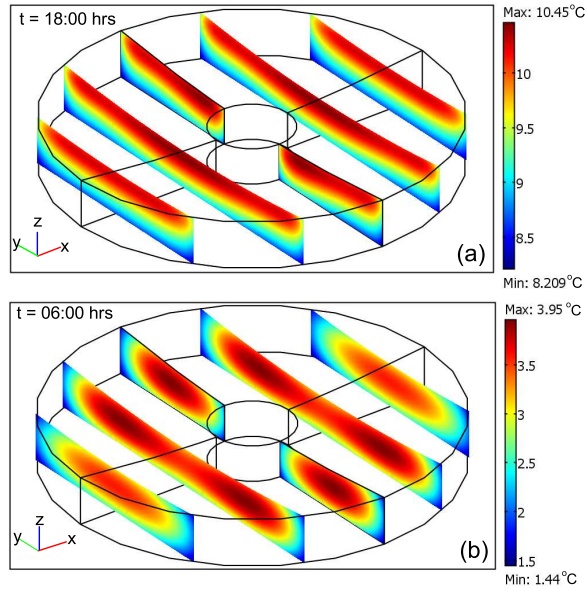}}
  \caption{Nonuniform temperature distribution inside the Zerodur mirror (a) due to solar heating during the day and (b) convective and radiative cooling during the night.}\label{fig}
\end{figure}
\subsection{Nonuniform thermal response}
Temperature of the mirror rises during the day-time observations when the heat produced at the front surface is conducted to the mirror interiors. The mirror remains significantly warmer during the night as the heat is slowly diffused now from mirror interiors to boundaries. On an average, the heating rate of the mirror is about 200W during the day-time. Considering the $10\%$ absorption, the total heat deposited in the mirror during every 12 hours run is $\sim 8.6\times10^6$~J.  As evident from the sliced plane visualization of the mirror volume shown in Fig. 3, the heat distribution within the Zerodur mirror is highly nonuniform.  The low thermal conductivities of Zerodur is primarily responsible for creating axial and radial temperature nonuniformities inside the mirror volume. The shifting temperature gradients can be inferred from the fact that a predominant heat concentration is observed in the upper half of the mirror during the day as in Fig. 3(a), while the overnight cooling effectively lowers the temperature leaving the core a little hotter than the rest of the mirror. But from the seeing perspective, it is the \emph{temperature difference} (hence forth denoted by $\Delta T$) between the mirror and the ambient air which is of a greater concern. The computed variations in $\Delta T$ for three reference points in Zerodur and SiC mirrors are shown in Fig. 4. For Zerodur, the temperature difference between the reflecting surface and ambient could change from 1$-5^\circ$C within 24 hours but it is always above the ambient. Therefore, to minimize the seeing effects, the front surface of the mirror has to be cooled continuously during the telescope operation. The middle and bottom section of the mirror can have temperature less than ambient during the day. The thermal lag between different parts of Zerodur material is because of its low thermal conductivity.

\begin{figure}[!h]
\centerline{\includegraphics[scale=0.24]{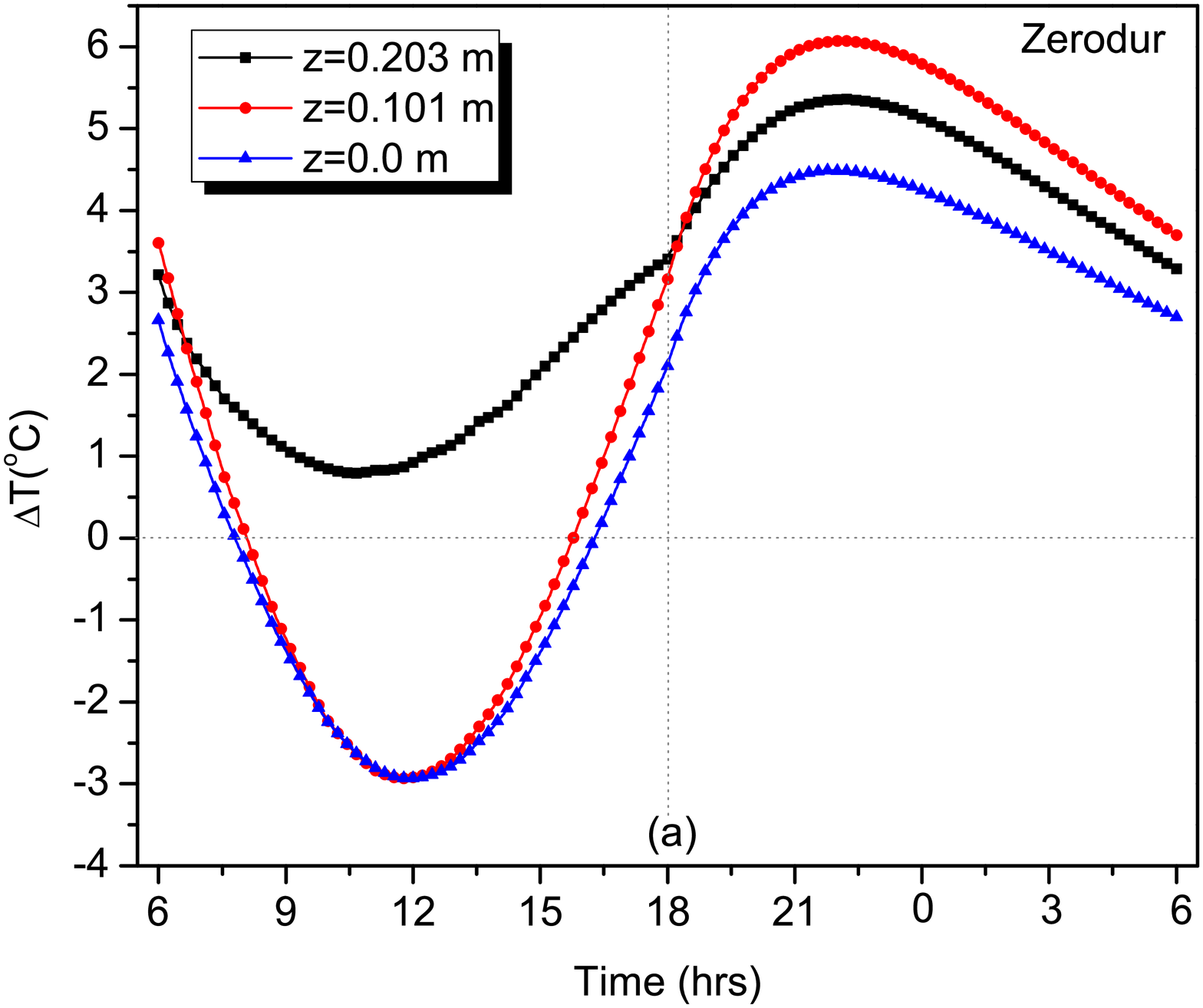} \includegraphics[scale=0.25]{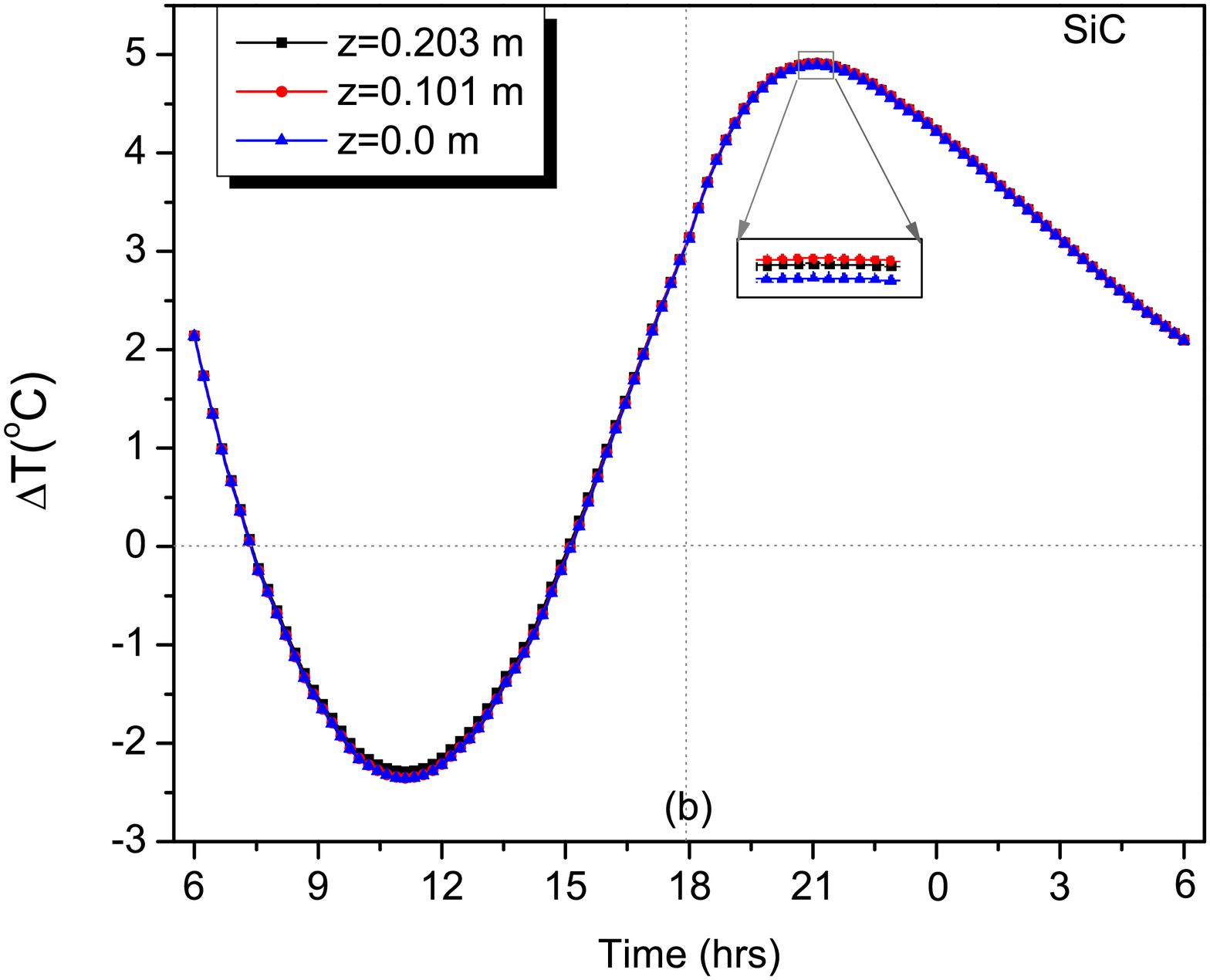}}
  \caption{Differential temperature variations evaluated for a reference point on top, middle and bottom of (a) Zerodur (b) SiC mirror. The ambient temperature range is $0-10^\circ$C.}\label{fig}
\end{figure}
In comparison to Zerodur, SiC has high thermal conductivity. Any heat injected into the mirror is quickly diffused into the entire substrate. As a result the entire SiC mirror block attains uniform temperature above ambient. For example, all three curves in Fig. 4(b) nearly overlap and there is  barely a noticeable difference ($<0.03^\circ$C) in their temperature.  Also the $\Delta T$ for the entire SiC remains below the zero difference line between 7:00-15:00 hrs. Even though the mirror is warmer than the ambient air by about $2^\circ$C at the beginning, it does not accumulate enough energy to raise the temperature of the entire block beyond ambient air until about 15:00 hrs. Under such conditions a mild heating of the front surface of SiC mirror during the early morning hours is necessary to bring its temperature close to the ambient.

\subsection{Radial and axial temperature gradients}
Figure (5) shows the day-time and night-time temperature difference of a radial line ($r\!=\!0.21$m to $r\!=\!1$m) along the front surface of the Zerodur and SiC mirrors. Temperature evolution is recorded for every two hours interval. Unlike SiC, the shape of the temperature curves shown in Fig. 5(a) and Fig. 5(b) for Zerodur mirror is highly distinct for the heating and cooling cycles. The gradual onset of solar flux in the morning hours and slow and complex heat diffusion across the substrate makes the temperature gradients highly nonuniform. The temperature gradients are particularly more pronounced closer to mirror edges. Further, as seen from Fig. 5(b), the predominant heat loss from sides during the nigh makes the temperature gradients even more steeper and smoother towards to mirror edges. The numerical calculations for the maximum temperature difference ($\Delta T_{\textrm{max}}$) existing between the mirror surface and ambient air at different times are recoded in the Table \ref{tab3}.

\begin{figure}[!h]
\begin{tabular}{cc}
\centerline{\includegraphics[scale=0.26]{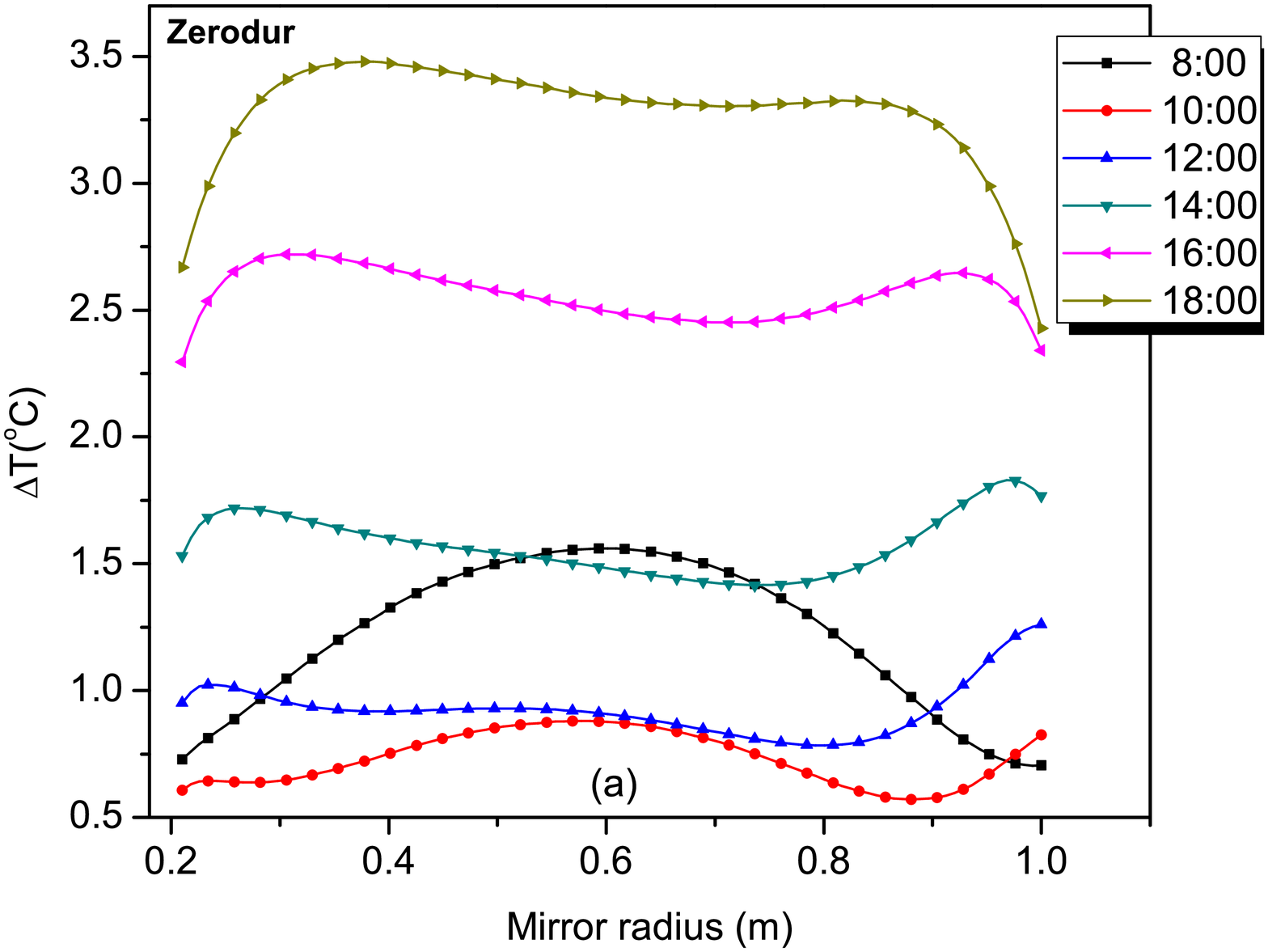} \includegraphics[scale=0.24]{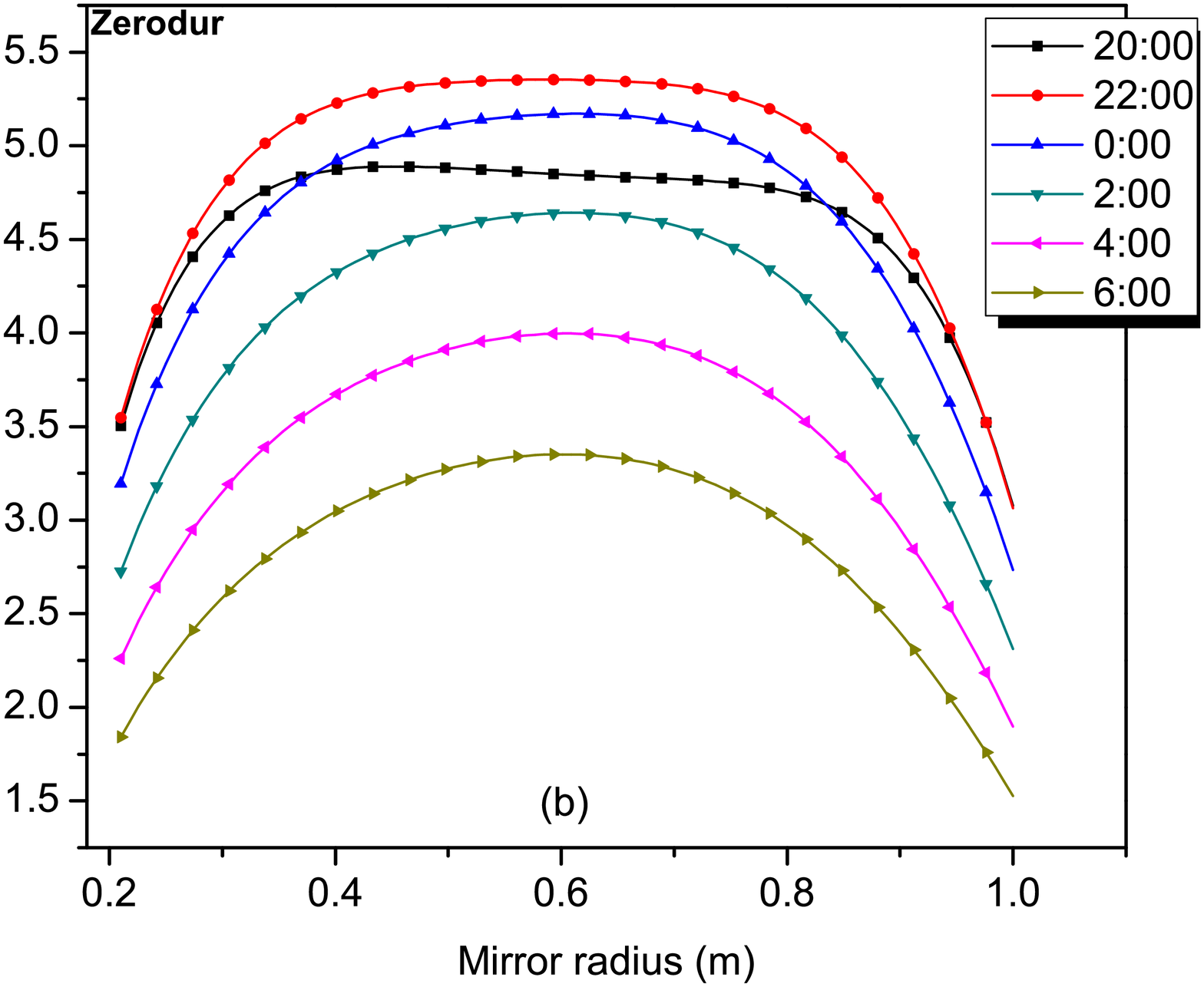}}\\
\centerline{\includegraphics[scale=0.26]{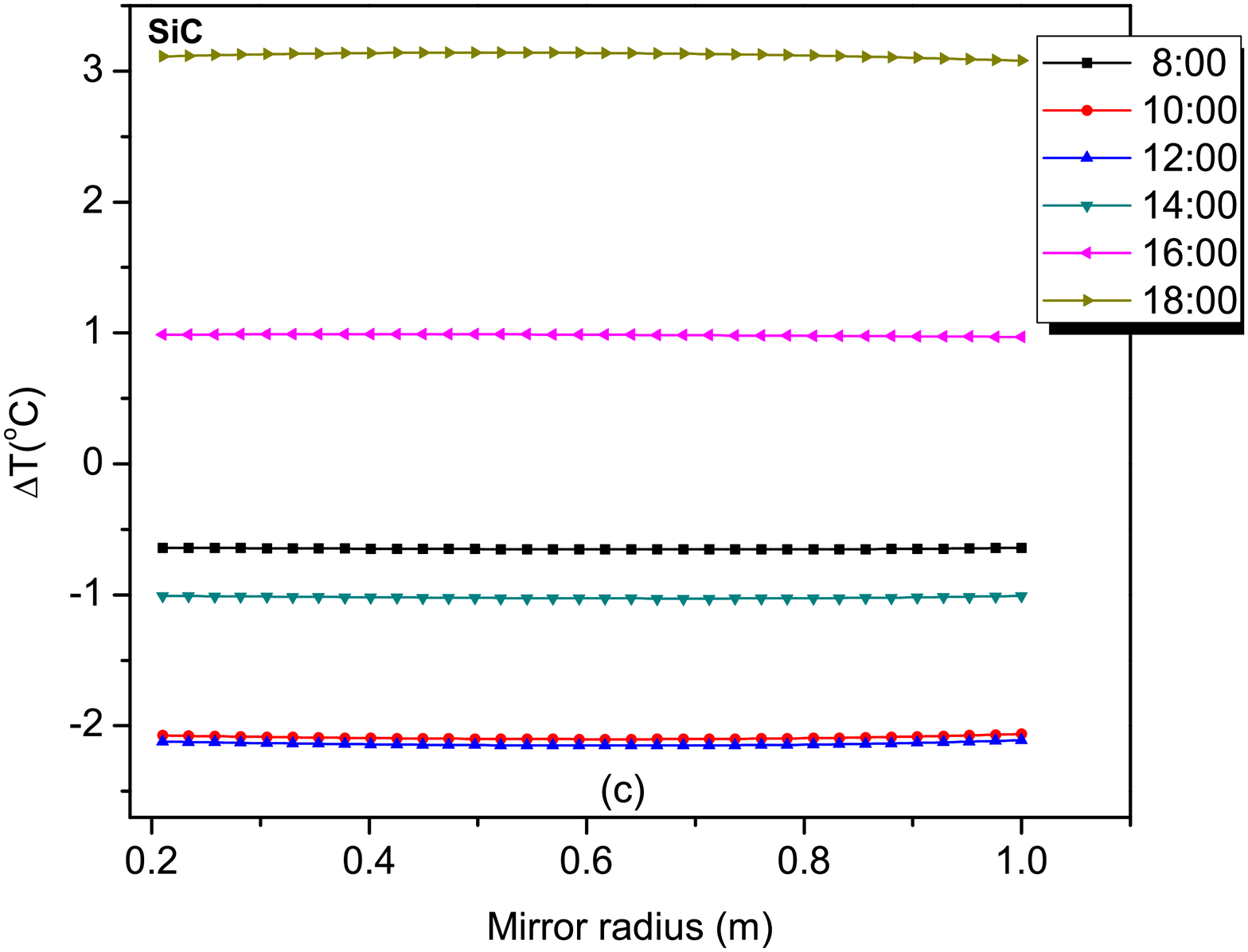} \includegraphics[scale=0.24]{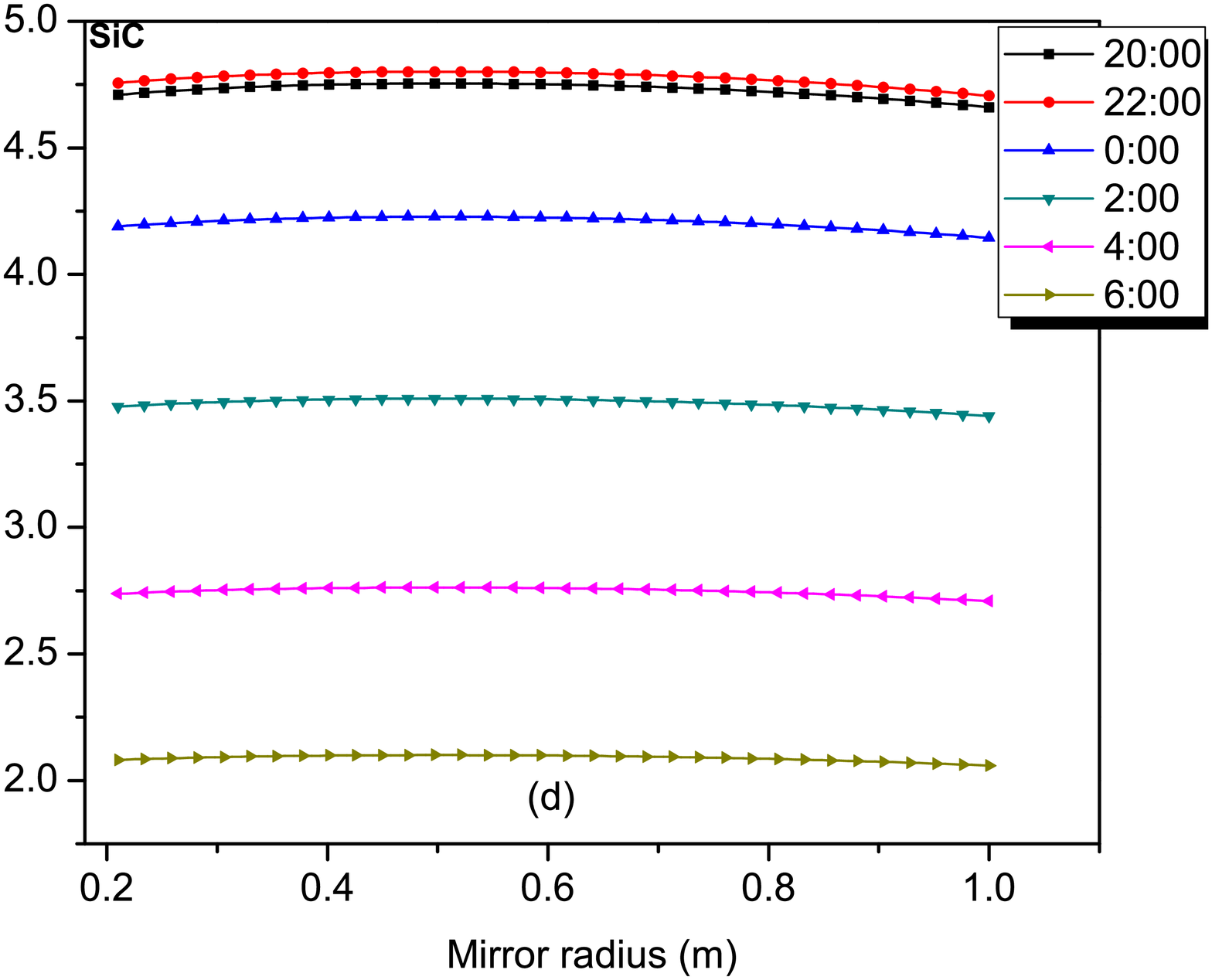}}
  \end{tabular}
  \caption{Radial temperature difference of the front mirror surface during day-time and night-time for (a)--(b) Zerodur mirror and (c)--(d) for SiC mirror. The ambient temperature range is $0-10^\circ$C. }\label{fig}
\end{figure}
Mirror cooling by natural convection is extremely inefficient.  So regardless of the material, the temperature difference between the mirror surface and surroundings tend to be predominantly high at night, making external cooling absolutely necessary.
\begin{table}[h]
  \centering  \caption {Excess mirror surface heating for Zerodur(SiC) mirror under different ambient temperature conditions.}   {\small
\begin{tabular}{|l|l|l|l|l|}
  \hline
Time &  \multicolumn{4}{|c|}{Ambient Temperature Range} \\  \cline{2-5}
(hrs.)  &  {$0^\circ\!\!-\!5^\circ$C} & {$0^\circ\!\!-\!10^\circ$C}& {$0^\circ\!\!-\!20^\circ$C} & {-$10^\circ\!\!-\!0^\circ$C} \\
&  $\Delta T_{\textrm{max}}$($^\circ$C) &    $\Delta T_{\textrm{max}}$($^\circ$C) &   $\Delta T_{\textrm{max}}$($^\circ$C) &   $\Delta T_{\textrm{max}}$($^\circ$C)  \\ \hline
 08:00  &  1.39(-0.17)  &  1.56(-0.64) &  0.83(-1.81)  &  -0.74(-2.42)  \\
 10:00  &  2.16(-0.50)  &  0.88(-2.06) &  -1.67(-5.58) &  -0.31(-3.46)  \\
 12:00  &  3.03(0.04)   &  1.26(-2.11) &  -1.70(-6.61) &  0.54(-3.04)  \\
 14:00  &  3.65(0.89)   &  1.83(-1.01) &  -0.59(-5.00) &  1.43(-1.68)  \\
 16:00  &  3.91(1.82)   &  2.72(0.99)  &  1.15(-1.26) &  2.17(0.33)  \\
 18:00  &  3.42(2.63)   &  3.48(3.14)  &  3.71(3.60)  &  2.81(2.62)  \\
 20:00  &  1.73(1.22)   &  4.89(4.75)  &  7.06(7.62)  &  4.29(4.39)  \\
 22:00  &  2.55(1.89)   &  5.35(4.80)  &  8.38(8.28)  &  4.84(4.54)  \\
 00:00  &  2.75(1.95)   &  5.17(4.23)  &  8.44(7.51)  &  4.79(4.04)  \\
 02:00  &  2.61(1.72)   &  4.64(3.51)  &  7.78(6.33)  &  4.36(3.38)  \\
 04:00  &  2.31(1.41)   &  4.00(2.76)  &  6.81(5.04)  &  3.80(2.67)  \\
 06:00  &  1.98(1.11)   &  3.35(2.10)  &  5.76(3.86)  &  3.21(2.03)  \\ \hline
  \hline
\end{tabular}}\label{tab3}
\end{table}

To investigate the temperature gradients along the mirror thickness, we chose the axial line ($z\!=\!0$m to $z\!=\!0.203$m) connecting the three reference points shown in Fig. 1. Again, the presence of strong and nonuniform axial gradients, particularly around the middle of the day and towards the midnight, is clearly evident for Zerodur in Fig. 6(a-b). For SiC the net temperature difference varies significantly. However, the apparent lack of axial and radial temperature gradients  can be seen clearly from Fig. 5(c-d) and Fig. 6 (c-d).

The radial temperature excess $\delta T_{\textrm{rad}}$ and axial temperature excess  $\delta T_{\textrm{axi}}$, which we define as the difference in maximum and minimum temperature occurring along the radial and axial directions at a particular instant of time, is computed for the Zerodur and SiC mirror at different ambient temperature range. These results shown in  Table \ref{tab4} indicate that the axial gradients are more severe during the day-time while the radial gradient along the front surface tend to dominate at night. A general conclusion which follows from a relative comparison of data in Table \ref{tab3} and Table \ref{tab4} is that day-time heating leads to `worse seeing' effects at night. Therefore, a solar telescope mirror cannot be readily adapted to carry out the intended astronomical observations at night unless the temperature of the mirror is actively controlled.
\begin{figure}[h]
\begin{tabular}{cc}
\centerline{\includegraphics[scale=0.25]{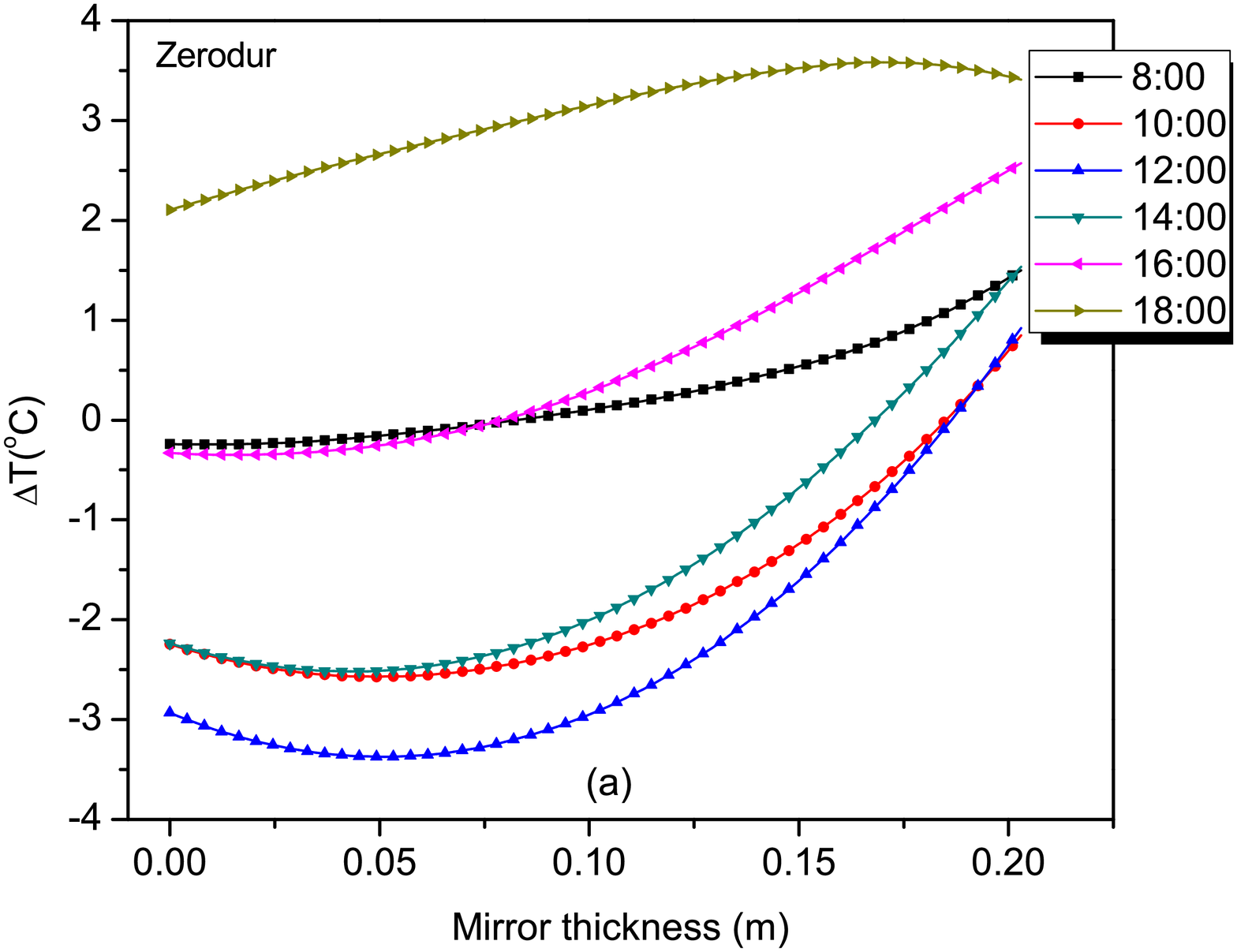} \includegraphics[scale=0.26]{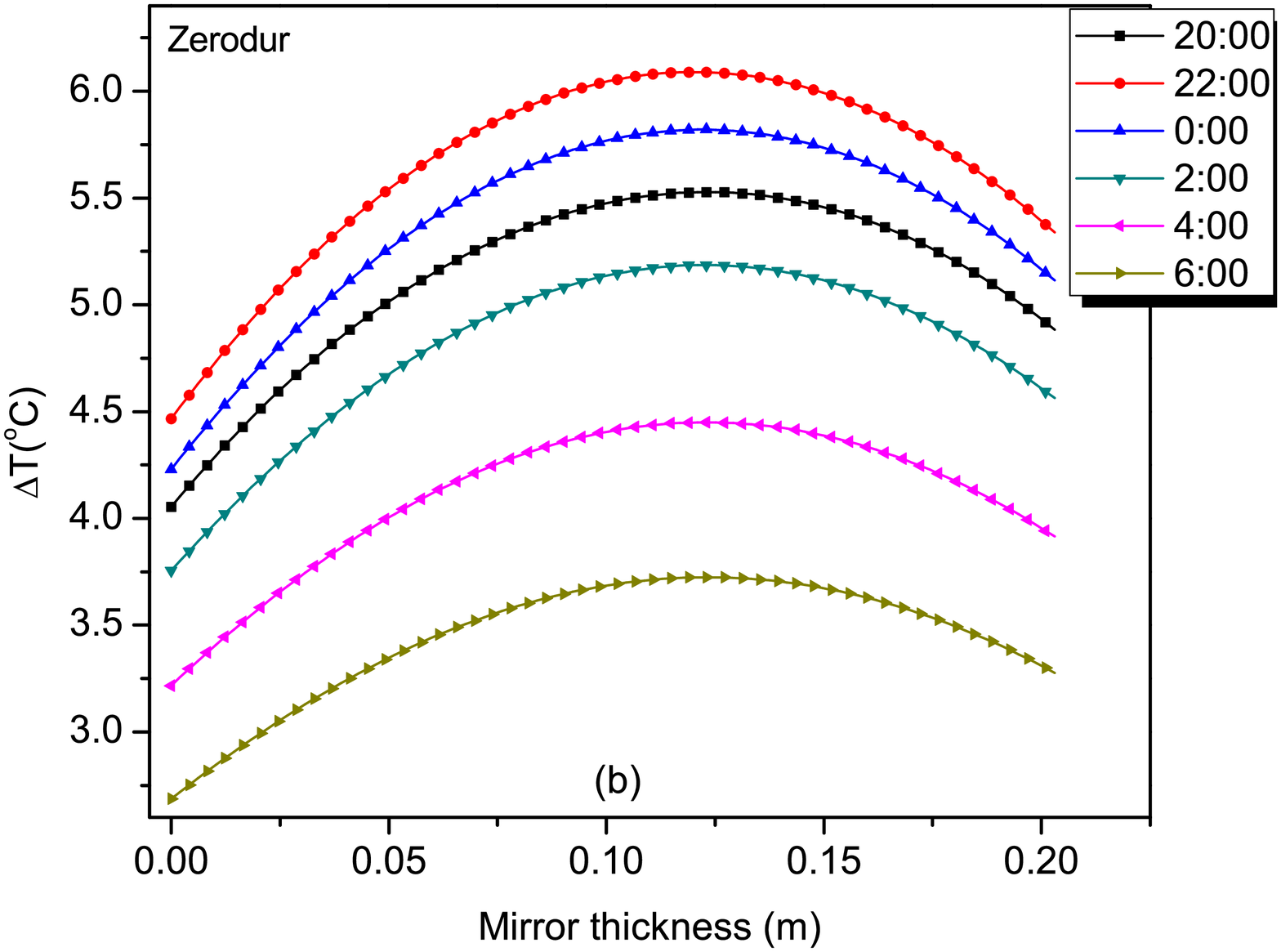}}\\
\centerline{\includegraphics[scale=0.25]{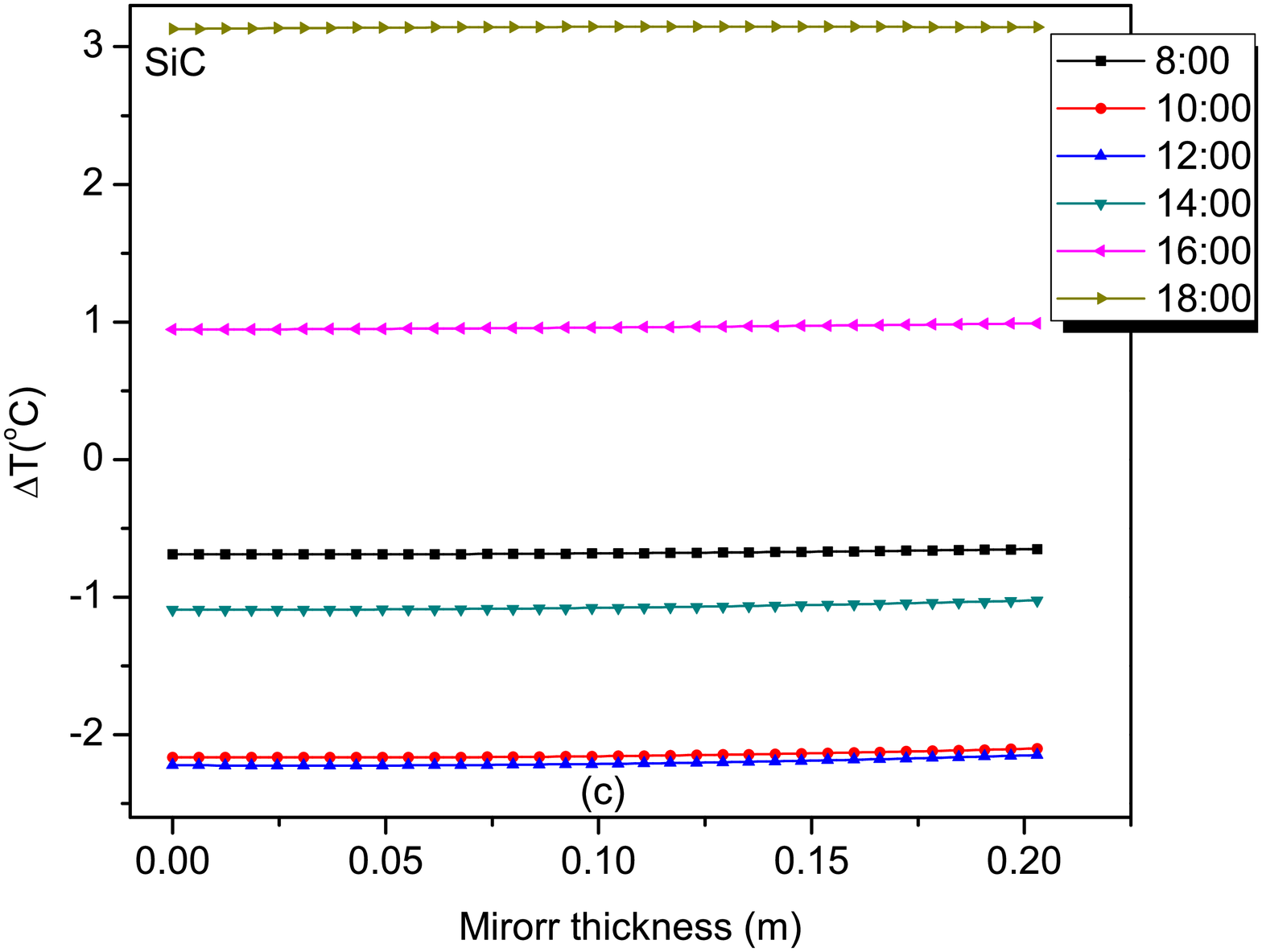} \includegraphics[scale=0.26]{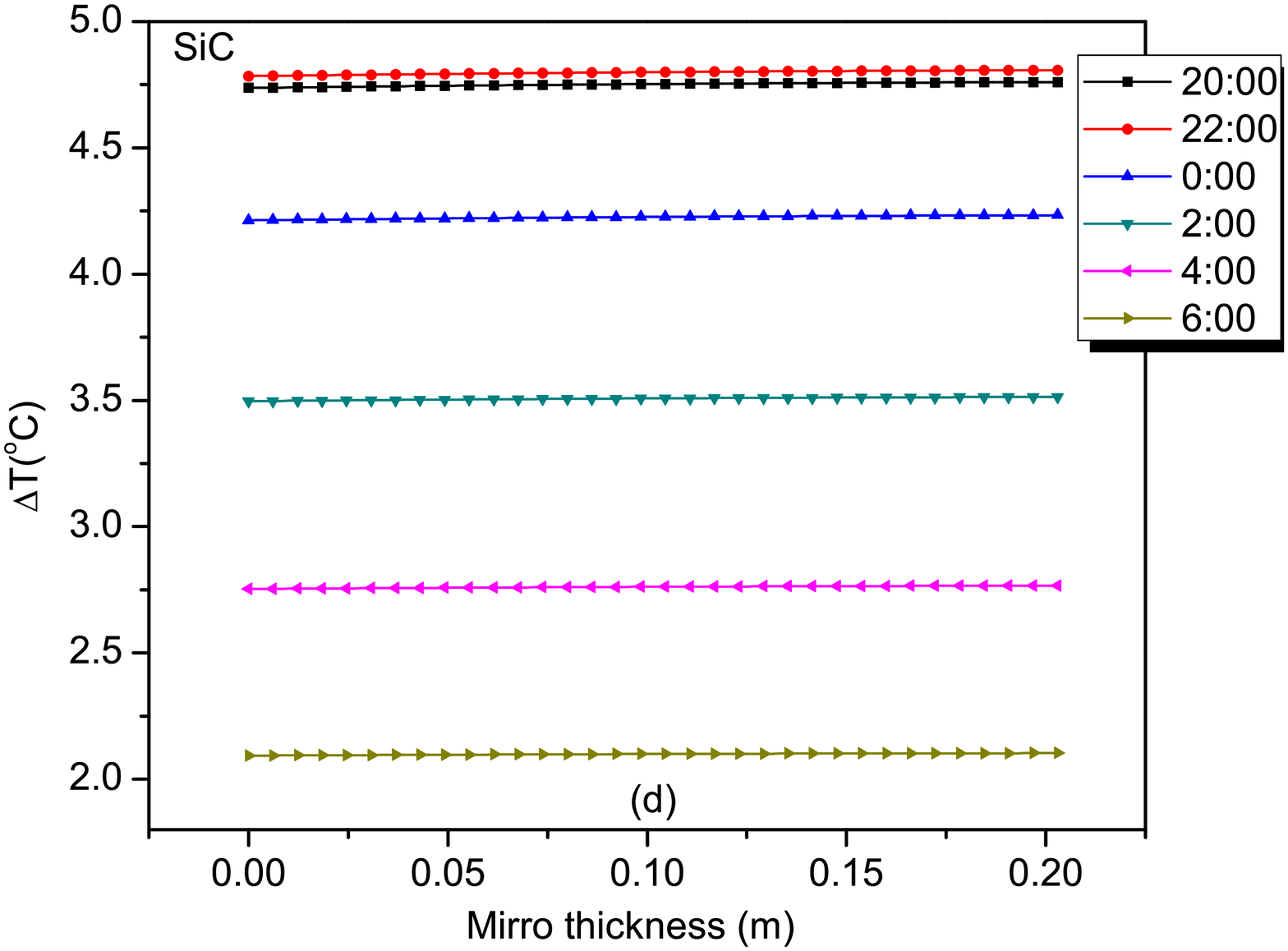}}
  \end{tabular}
  \caption{Axial temperature difference along the thickness of the mirror blank during day and night-time for (a)--(b) Zerodur material, and (c)--(d) for SiC. The ambient temperature range is $0-10^\circ$C.}\label{fig}
\end{figure}

\begin{table}[!h]
\caption {Numerically computed radial ($\delta T_{\textrm{rad}}$) temperature excess  and axial temperature excess ($\delta T_{\textrm{axi}}$)  for the Zerodur(SiC) mirror at different ambient temperature range.} \label{tab4}
  \centering   {\tiny
\begin{tabular}{|l|ll|ll|ll|ll|}
  \hline
  Time &  \multicolumn{8}{|c|}{Ambient Temperature Range} \\ \cline{2-9}
  (hrs)  &  \multicolumn{2}{c|}{$0^\circ\!\!-\!5^\circ$C} & \multicolumn{2}{c|}{$0^\circ\!\!-\!10^\circ$C}& \multicolumn{2}{c|}{$0^\circ\!\!-\!20^\circ$C} & \multicolumn{2}{c|}{-$10^\circ\!\!-\!0^\circ$C} \\
    &  $\delta T_{\textrm{rad}}$($^\circ$C) &   $\delta T_{\textrm{axi}}$($^\circ$C) &   $\delta T_{\textrm{rad}}$($^\circ$C) &   $\delta T_{\textrm{axi}}$($^\circ$C) &   $\delta T_{\textrm{rad}}$($^\circ$C) &   $\delta T_{\textrm{axi}}$($^\circ$C) &   $\delta T_{\textrm{rad}}$($^\circ$C) &   $\delta T_{\textrm{axi}}$($^\circ$C) \\ \hline
 08:00  &  0.44(0.01)  &  1.64(0.04) &  0.86(0.01) &  1.74(0.04) &  1.26(0.03) &  1.84(0.02) &  0.62(0.04) &   1.72(0.04) \\
 10:00  &  0.42(0.02)  &  3.34(0.07) &  0.31(0.04) &  3.42(0.07) &  1.27(0.11) &  3.71(0.06) &  1.23(0.06) &   3.60(0.07) \\
 12:00  &  0.53(0.02)  &  3.58(0.08) &  0.48(0.04) &  4.30(0.08) &  2.37(0.14) &  4.74(0.08) &  1.51(0.06) &   4.58(0.08) \\
 14:00  &  0.71(0.02)  &  4.15(0.07) &  0.41(0.02) &  4.05(0.07) &  2.61(0.10) &  4.44(0.06) &  1.29(0.03) &   4.34(0.07) \\
 16:00  &  0.94(0.04)  &  3.19(0.05) &  0.42(0.02) &  2.92(0.05) &  1.72(0.03) &  2.86(0.04) &  0.76(0.01) &   2.96(0.04) \\
 18:00  &  1.23(0.05)  &  1.57(0.01) &  1.05(0.06) &  1.48(0.02) &  0.78(0.07) &  1.22(0.02) &  0.66(0.05) &   1.25(0.01) \\
 20:00  &  1.54(0.06)  &  1.15(0.02) &  1.81(0.09) &  1.47(0.02) &  2.11(0.15) &  1.97(0.04) &  1.41(0.08) &  1.22(0.02) \\
 22:00  &  1.67(0.06)  &  1.10(0.01) &  2.23(0.10) &  1.62(0.02) &  3.16(0.17) &  2.48(0.04) &  1.91(0.08) &  1.41(0.02) \\
 00:00  &  1.65(0.05)  &  1.03(0.01) &  2.44(0.08) &  1.59(0.02) &  3.65(0.15) &  2.54(0.04) &  2.14(0.07) &  1.42(0.02) \\
 02:00  &  1.51(0.04)  &  0.90(0.01) &  2.33(0.07) &  1.43(0.02) &  3.73(0.13) &  2.50(0.03) &  2.10(0.06) &  1.29(0.02) \\
 04:00  &  1.33(0.03)  &  0.13(0.01) &  2.10(0.05) &  1.23(0.01) &  3.48(0.10) &  2.11(0.03) &  1.93(0.05) &  1.31(0.01) \\
 06:00  &  1.14(0.02)  &  0.63(0.01) &  1.83(0.04) &  1.04(0.01) &  3.08(0.08) &  1.78(0.02) &  1.70(0.04) &  0.96(0.01) \\ \hline \hline
  \end{tabular} }
\end{table}
\begin{figure}[!h]
\centerline{\includegraphics[scale=0.27]{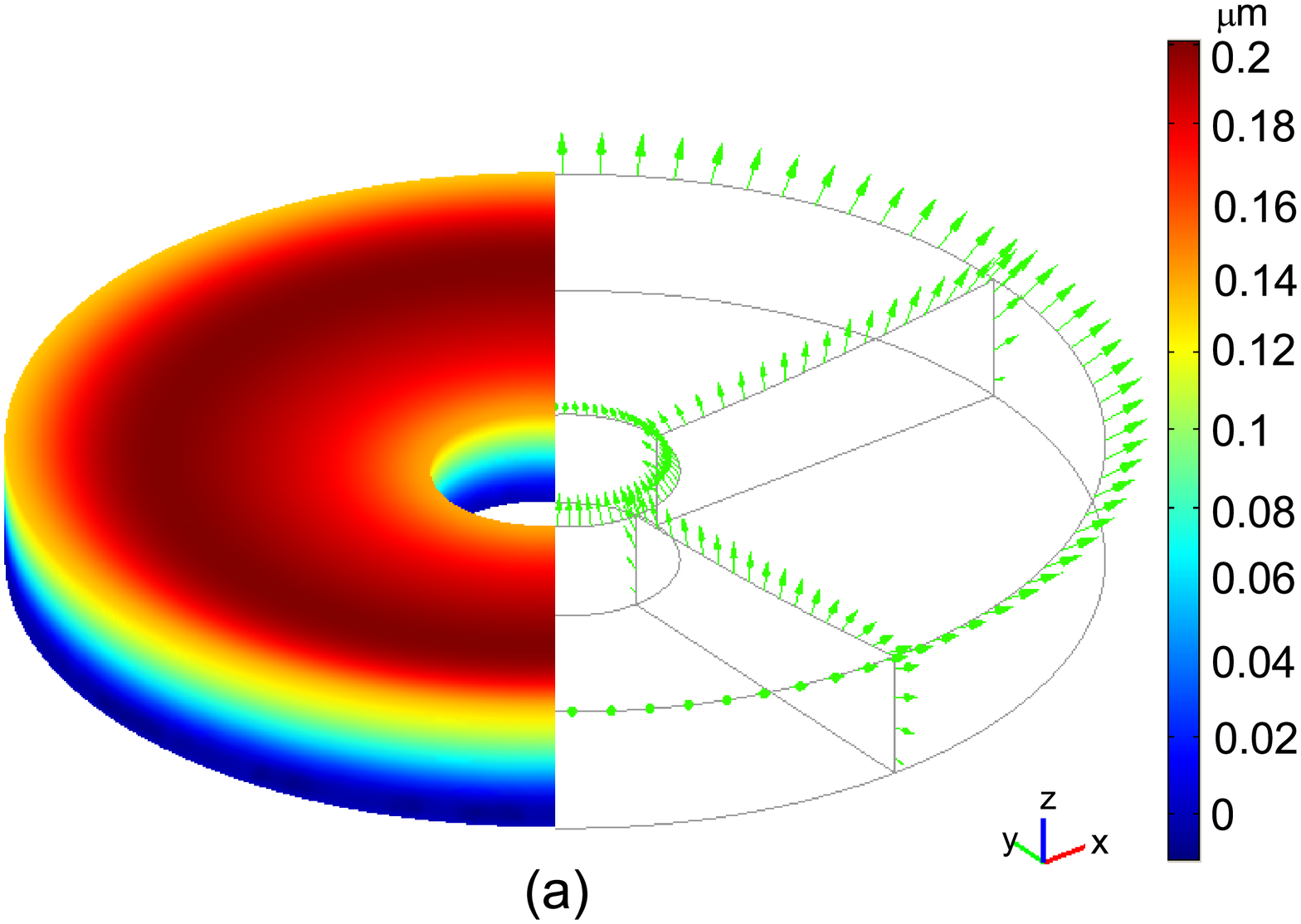} \includegraphics[scale=0.27]{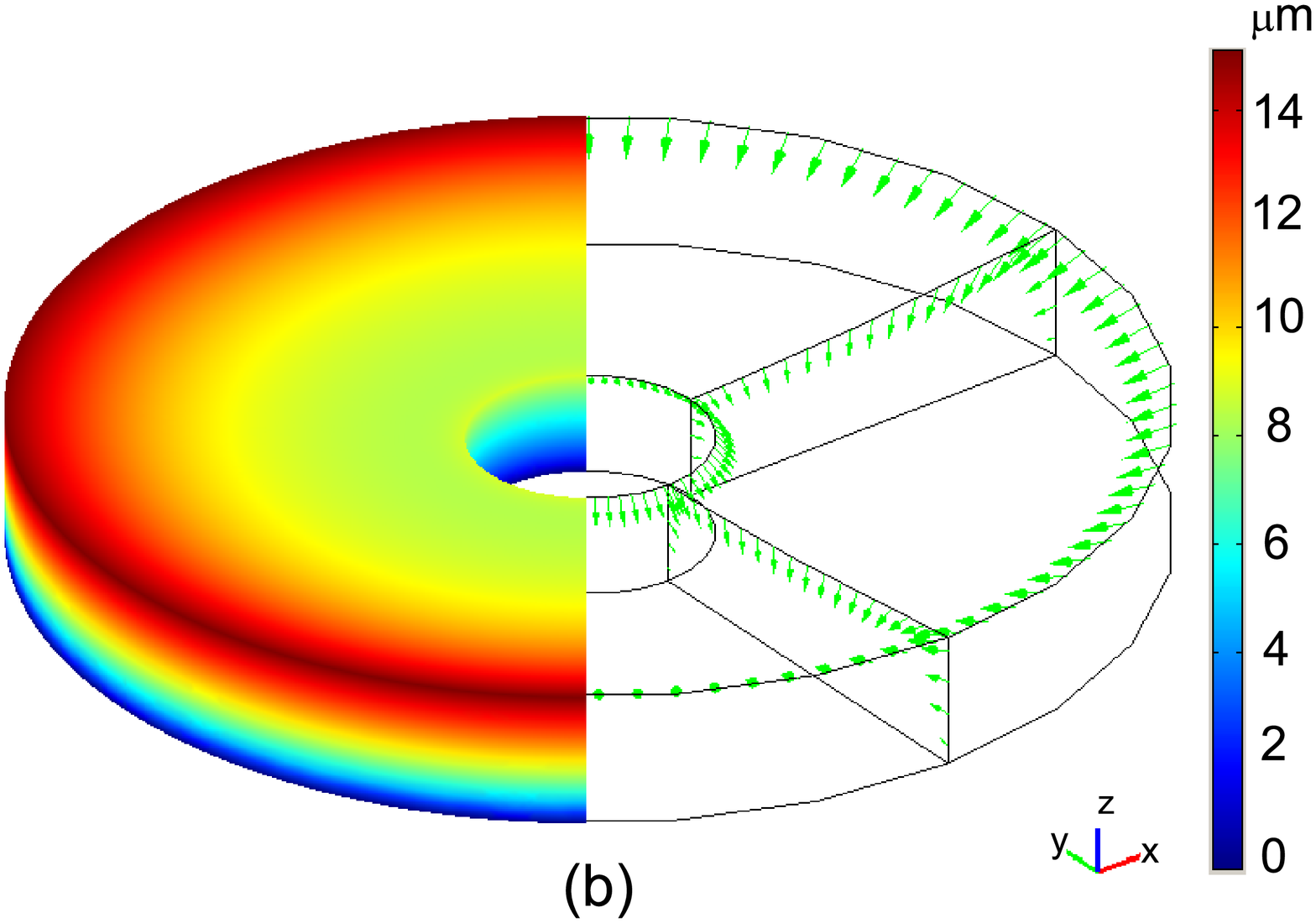}}
  \caption{A snapshot of the surface distortion of (a) Zerodur mirror (b) SiC mirror for ($0-10^\circ$C) ambient temperature change. The direction of the displacement vectors at the mirror surface is denoted by arrows shown in green.}\label{fig}
\end{figure}

\subsection{Thermal deformation}
The conductive flow of heat energy inside a solid material establishes a temperature field which induces thermal stresses  producing a volumetric changes in the material body. For an isotropic case, the thermal strain $ \varepsilon$ in an elastic solid depends on the temperature $T$ and the stress-free reference temperature $T_{\textrm{ref}}$ which is given by:
\begin{equation}\label{eqs}
    \varepsilon=\alpha \cdot (T-T_{\textrm{ref}})
\end{equation}
The constitutive strain-displacement relationship for an uncoupled thermoelastic problem can be expressed for normal strain as \citep{sad04}:
\begin{equation}
 \varepsilon_{x} = \frac{\partial u}{\partial x} ;\;\; \varepsilon_{y}=  \frac{\partial v}{\partial y}; \;\;\;  \varepsilon_{z} =  \frac{\partial w}{\partial z}
  \end{equation} and for shear strain as:
  \begin{eqnarray}
     \varepsilon_{xy} &=& \frac{1}{2}\left(\frac{\partial u}{\partial y}+\frac{\partial v}{\partial x} \right)   \\
    \varepsilon_{yz} &=& \frac{1}{2}\left(\frac{\partial v}{\partial z}+\frac{\partial w}{\partial y}\right)   \\
   \varepsilon_{zx} &=& \frac{1}{2}\left(\frac{\partial w}{\partial x}+\frac{\partial u}{\partial z}\right)
  \end{eqnarray} where $u$, $v$ and $w$ are the displacement components in $x$, $y$ and $z$ directions, respectively.
The FEM numerical implementation of Eq.(7)-Eq.(11) in the COMSOL Multiphysics software is based on the principle of virtual work done. The departure of  reflecting surface of the telescope mirror from its nominal shape is computed by calculating the displacement $p=\sqrt{u^2+v^2+w^2}$ of the surface node positions  from the undeformed geometry.  The discrete sag data $ds$ calculated using $ds=w-z^\prime\sqrt{u^2+v^2}$ is particularly useful for representing the surface irregularities in terms of Zernike polynomials \citep{doy02}.  For now, we only present a measure of displacement errors across the mirror surface computed in terms of the $p_{rms}\!=\!\int p\,dA /A$ and $w_{rms}\!=\!\int w\,dA /A$, where $A$ is the area of the reflecting surface of the mirror. The optical aberrations are proportional to the rms amplitude, $p_{rms}$ and $w_{rms}$. Though it does not tell us directly the the shape  of the irregularities.

Fig. 7 shows a graphical snapshot of the front surface distortions of Zerodur and SiC mirror obtained by solving  structural FEM model. The 3D thermal profiles were supplied as input to the FEM model when the temperature of the mirrors had reached maximum value during the day-time heating.  In this case, the Zerodur and SiC mirror has reached the maximum surface temperature about $12.2^\circ$C and $10.3^\circ$C, at about 14:40 hrs and 16:30 hrs, respectively. The large thermal deformations along the mirror edges are greatly due to the curved shape of the mirror surface.
\begin{figure}[h]
\centerline{\includegraphics[scale=0.25]{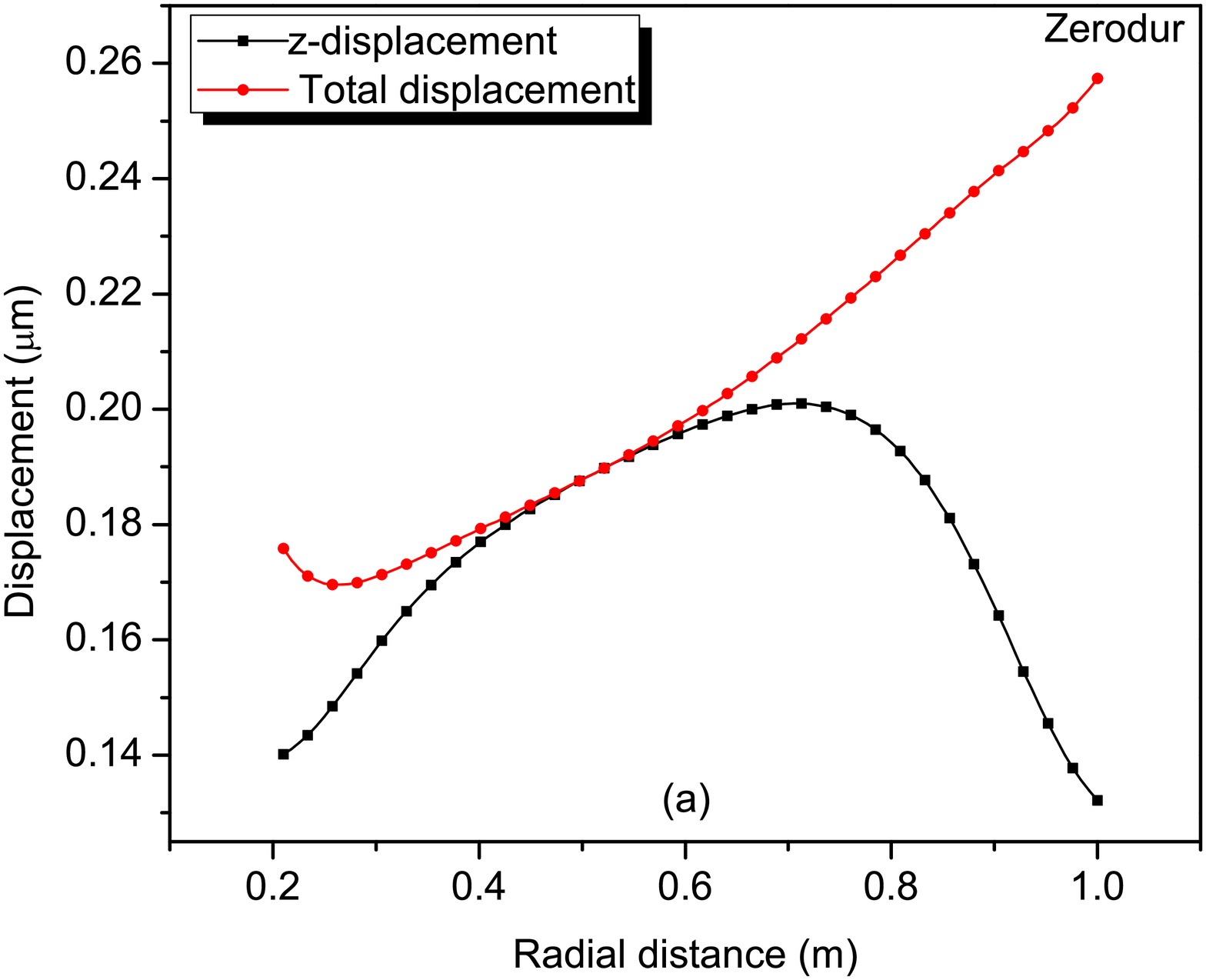} \includegraphics[scale=0.25]{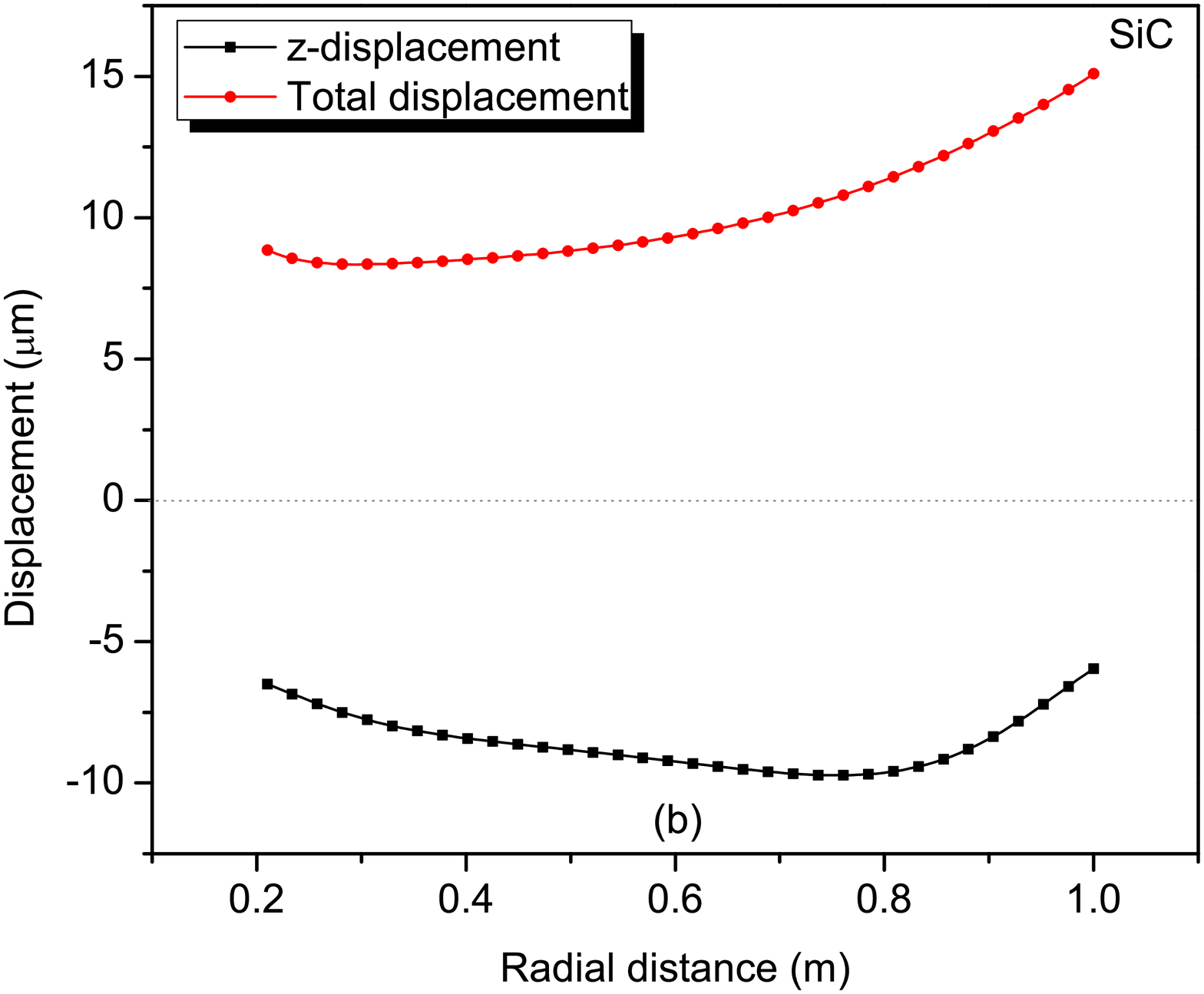}}
  \caption{Calculated displacement along the top surface of (a) Zerodur (b) SiC mirror.}\label{fig}
\end{figure}

As discussed in  previous section, there are strong thermal gradients within the Zerodur mirror and the resulting surface distortions can also be highly irregular. However, due to its extremely low CTE, the magnitude of the thermal deformations remains incomparably low. Also, the value of CTE is negative over the operating temperature range. The displacement vectors as seen in the Fig. 7(a), are all pointing outward. So the z-displacement and sag, however little they may be, are all positive. Though, the near-zero CTE of zerodur is useful to keep the thermal deformation at negligible level, the wavefront errors caused by highly nonuniform surface deformations, cannot be corrected by simple focus shift.

For SiC, the displacement vectors are pointing into the mirror. The sag and the z-displacement is, therefore, clearly negative. More importantly, these surface perturbations are not uniform over the mirror surface. A typical line profile for the total displacement and z-displacement occurring at the peak mirror temperature during the day for Zerodur and SiC are shown in Fig. 8. Even though the temperature distribution inside the SiC mirror is highly uniform, the total displacement is increasing radially outward from $\sim\!9\mu$ to  $\sim\!15\mu$ in a quadric manner. This is not unexpected since the original mirror shape itself is parabolic. For a mirror blank, the lack of temperature gradients in SiC is one of the most distinguishing and a highly desirable feature. An uniform temperature rise in SiC mirror does not produce significant higher order aberrations. It would only change its global radius of curvature, causing the defocus error in the final optical image. The defocus error can be easily corrected by moving the secondary mirror. A smooth quadric variations in the surface displacement shown in Fig. 8(b) would support this conclusion.

The detailed calculations of $p_{rms}$ and $w_{rms}$ at different temperature range for two mirror materials are summarized in Table \ref{tab5}.  In the FEA structural model, the reference temperature $T_{\textrm{ref}}$ for mirror blanks is assumed to be $20^\circ$C. The magnitude of thermal distortions would depends on how much the mirror temperature deviates from the reference temperature. The calculations shown in Table \ref{tab5} are therefore, consistent with expected response of the materials under specified temperature range.
It is also important to note that thermal distortions would vary with the varying temperature conditions which requires active or adaptive optics support for corrections.

\begin{table}[!h]
  \centering   \caption{The numerical values of the thermal deformation calculated for the front surface of the Zerodur and SiC mirror at different ambient temperature range.}\label{tab5}
  \begin{tabular}{|c|c|c|c|c|} \hline
  Temp. range &  \multicolumn{2}{|c|}{$p_{rms}$ ($\mu$m)} & \multicolumn{2}{|c|} {$w_{rms}$($\mu$m)} \\
   & Zerodur & SiC & Zerodur & SiC\\ \hline
  -10--0$^\circ$C & 0.625& 331.28& 0.506 & -261.76\\
   0--5$^\circ $C &0.312 & 15.33& 0.260 &-12.13\\
   0--10$^\circ $C &0.216 &10.95 &0.179 &-8.66 \\
   0--15$^\circ $C &0.135& 6.95& 0.112& -5.50\\
   0--20$^\circ $C & 0.072 & 0.282 & 0.058& -1.81\\  \hline \hline
  \end{tabular}
  \end{table}

\section{Conclusion}
The primary mirror heating caused by the light absorption is a complex and challenging problem. The thermal performance of a solar telescope mirror needs to be accurately predicted under real observing conditions. This is essential to design an efficient and durable temperature control system devised to mitigate the detrimental effects of excessive heating during the day. We have outlined an approach to study the thermal and structural response of  a primary  mirror under varying observing conditions. In the  FEM model for a 2m class primary mirror, the location dependent solar flux and a simple physics based heating and cooling model of the ambient air temperature was incorporated. The spatial and temporal evolution of temperature field inside two well known materials for optical telescope, Silicon Carbide (SiC) and Zerodur, were examined using 3D numerical simulations.

The low thermal conductivity of Zerodur mirror gives rise to strong radial and axial temperature gradients that are quite distinct for the day-time heating and night-time cooling. Heat loss by free convection is very slow so the mirror retain significant heat during the night. The thermal response of the SiC mirror is significantly  different from the Zerodur. The temperature within the SiC mirror substrate equilibrates rather quickly due to high thermal conductivity. The absence of thermal gradients  and the advantage of high thermal conductivity of SiC cannot be favorably leveraged without some temperature regulation by external means. Thermal distortions of the mirror were analyzed using the structural FEM model. High surface distortion seems to result if the operating temperature of the mirror deviates significantly from the nominal temperature of the material. In extremely low CTE materials, the mirror seeing ultimately limits the telescope performance. It is not the high temperature alone, but the relative incremental change in ambient air temperature and the mirror which contribute most to the `seeing effect'. Large temperature changes in ambient air and slow thermal response of the glass materials invariably results in bad seeing. In order to augment the scientific productivity of a solar telescope, there have been some suggestions from the astronomers about the possibility of utilizing the same telescope for limited, but useful night-time observations.  A site with minimum diurnal changes in day-night temperature (e.g. 0-5$^\circ$ C) would ideally suit for such dual purpose observations. We have only considered the thermal gradients and surface distortions in solid mirror which would be quite different for the lightweighted mirror made of same materials. The most conspicuous fallout of the different cell geometries and side walls structure is the appearance of thermal footprint in the surface distortions.  This would be the subject matter of future work.
\section{Acknowledgment} We thank Dr S. Chatterjee for several discussions and valuable suggestions that helped us to improve the form and content of the manuscripts.

\end{document}